\documentclass{article}

\usepackage{PRIMEarxiv}

\usepackage[utf8]{inputenc} 
\usepackage[T1]{fontenc}    
\usepackage{hyperref}       
\usepackage{url}            
\usepackage{booktabs}       
\usepackage{amsfonts}       
\usepackage{nicefrac}       
\usepackage{microtype}      
\usepackage{lipsum}
\usepackage{fancyhdr}       
\usepackage{graphicx}       
\graphicspath{{media/}}     
\usepackage{mathtools}  

\title{FD-Net: An Unsupervised Deep Forward-Distortion Model for Susceptibility Artifact Correction in EPI
}

\author{
  Abdallah Zaid Alkilani\\
  Department of Electrical and Electronics Engineering \& \\
  National Magnetic Resonance Research Center (UMRAM) \\
  Bilkent University \\
  Ankara, Turkey \\
  \texttt{alkilani@ee.bilkent.edu.tr} \\
   \And
  Tolga Çukur, Emine Ulku Saritas \\
  Department of Electrical and Electronics Engineering \&\\
  National Magnetic Resonance Research Center (UMRAM) \&\\
  Neuroscience Graduate Program \\
  Bilkent University \\
  Ankara, Turkey \\
  \texttt{\{cukur, saritas\}@ee.bilkent.edu.tr} \\
}

\begin{document}
\maketitle

\begin{abstract}
Recent learning-based correction approaches in EPI estimate a displacement field, unwarp the reversed-PE image pair with the estimated field, and average the unwarped pair to yield a corrected image.
Unsupervised learning in these unwarping-based methods is commonly attained via a similarity constraint between the unwarped images in reversed-PE directions, neglecting consistency to the acquired EPI images.
This work introduces an unsupervised deep-learning method for fast and effective correction of susceptibility artifacts in reversed phase-encode (PE) image pairs acquired with EPI.
FD-Net predicts both the susceptibility-induced displacement field and the underlying anatomically-correct image.
Unlike previous methods, FD-Net enforces the forward-distortions of the correct image in both PE directions to be consistent with the acquired reversed-PE image pair.
FD-Net further leverages a multiresolution architecture to maintain high local and global performance.
FD-Net performs competitively with a gold-standard reference method (TOPUP) in image quality, while enabling a leap in computational efficiency.
Furthermore, FD-Net outperforms recent unwarping-based methods for unsupervised correction in terms of both image and field quality.
The unsupervised FD-Net method introduces a deep forward-distortion approach to enable fast, high-fidelity correction of susceptibility artifacts in EPI by maintaining consistency to measured data.
Therefore, it holds great promise for improving the anatomical accuracy of EPI imaging.
\end{abstract}


\section{Introduction}\label{sec:intro}

Echo planar imaging (EPI)~\cite{Mansfield_1977} is the most commonly employed MRI sequence for diffusion-weighted imaging (DWI) and functional MRI (fMRI), due to its rapid k-space acquisition capability~\cite{Holdsworth_2008, Deichmann_2003}.
However, EPI is prone to susceptibility artifacts arising from B\textsubscript{0} field inhomogeneities, which are particularly prominent near tissue interfaces~\cite{Ludeke_1985}.
These artifacts manifest as intensity distortions from signal pileups/dropouts, and geometrical distortions due to compression/stretching of affected regions~\cite{Chang_1992}.
Severe artifacts can limit the clinical utility of EPI images.
Therefore, artifact correction is an essential step to ensure accuracy of downstream qualitative and quantitative assessments, especially at high field strengths~\cite{Tournier_2011,Jezzard_2012, Gallichan_2018}.

A leading framework for susceptibility-artifact correction uses images acquired in reversed phase-encoding (PE) directions to estimate the susceptibility-induced displacement field directly from the resulting blip-up (BU) and blip-down (BD) EPI images~\cite{Chang_1992, Andersson_2003, Holland_2010, Morgan_2004}.
An unwarping-based approach is commonly adopted for correction, where the reversed-PE images are nonlinearly transformed to alleviate artifacts based on the estimated displacement field.
Either voxel-wise field estimates~\cite{Ardekani_2005, Embleton_2010, Holland_2010}, or weighted combination of basis spatial maps across the field-of-view (FOV)~\cite{Andersson_2003} can be used.
Popular implementations of this framework include classical methods such as TOPUP from the FMRIB Software Library (FSL)~\cite{Smith_2004,Andersson_2003} and hyperelastic susceptibility correction of DTI data (HySCO) from the Statistical Parametric Mapping (SPM) toolbox~\cite{Ruthotto_2013,Modersitzki_2009}.
Since no additional data collection is needed beyond reversed-PE images, classical methods in the unwarping-based framework can offer notable benefits over measured-field-based, registration-based, or point spread function (PSF) based approaches in the literature~\cite{Graham_2017,Patzig_2021}.
Nonetheless, these classical methods are based on iterative optimization techniques that introduce substantial computational burden.

Deep neural networks have recently been considered as a powerful alternative for artifact correction that can maintain high computational efficiency~\cite{Ronneberger_2015}.
Previous studies in this domain have adopted the unwarping-based framework where reversed-PE images are first individually unwarped, and then combined to produce a final estimate.
In the absence of ground-truth for anatomically-correct images, network training has been performed via an unsupervised learning strategy that aims to maximize the similarity of unwarped images across the two PE directions~\cite{Duong_2020}.
Among previous learning-based methods, S-Net performs unwarping via bilinear interpolation and assesses the similarity between the corrected BU/BD images via a cross-modal loss~\cite{Duong_2020}.
Deepflow-Net instead performs unwarping via cubic interpolation and assesses the similarity between the corrected BU/BD images via a mean-squared error (MSE) loss~\cite{Zahneisen_2020}.
While promising results have been reported, these previous methods define an unsupervised loss function in the output domain of unwarped images, for which no ground-truth data are available.
Such lack of physical constraints in the loss function can cause suboptimal learning~\cite{Hammernik_2018, Aggarwal_2020}.
In turn, the network can produce low-fidelity images during inference, resulting in solutions that are notably inconsistent with the acquired reversed-PE images~\cite{Andersson_2014_bc}.

Here, we propose a novel deep network model (FD-Net) based on a forward-distortion approach for correcting EPI susceptibility artifacts in reversed-PE image pairs.
Unlike unwarping-based methods that average individually-corrected reversed-PE images, FD-Net predicts a single anatomically-corrected image along with a displacement field.
Unlike previous deep-learning methods, FD-Net directly incorporates physical constraints in the input domain where measurements are available.
Specifically, FD-Net forward-distorts the corrected image with the predicted field to reconstruct the reversed-PE image pair.
Unsupervised learning is then achieved by enforcing consistency of the reconstructed versus acquired reversed-PE images.
A multiresolution architecture is employed to maintain performance at both local and global scales.
Comprehensive demonstrations are performed to assess the quality of corrected images and field estimates on EPI data from the Human Connectome Project (HCP) database~\cite{VanEssen_2013}.
FD-Net performs competitively with the reference TOPUP method, while enabling a leap in computational efficiency; and it significantly outperforms competing deep-learning methods based on the unwarping framework.
These findings demonstrate the potential of FD-Net as a fast and effective method for susceptibility-artifact correction in EPI.


\section{Theory}\label{sec:theory}

    \subsection{Susceptibility Induced Distortions}\label{subsec:susceptibility_induced_distortions}
    
\begin{figure}[htbp]
    \centering
    \includegraphics[width=0.5\textwidth]{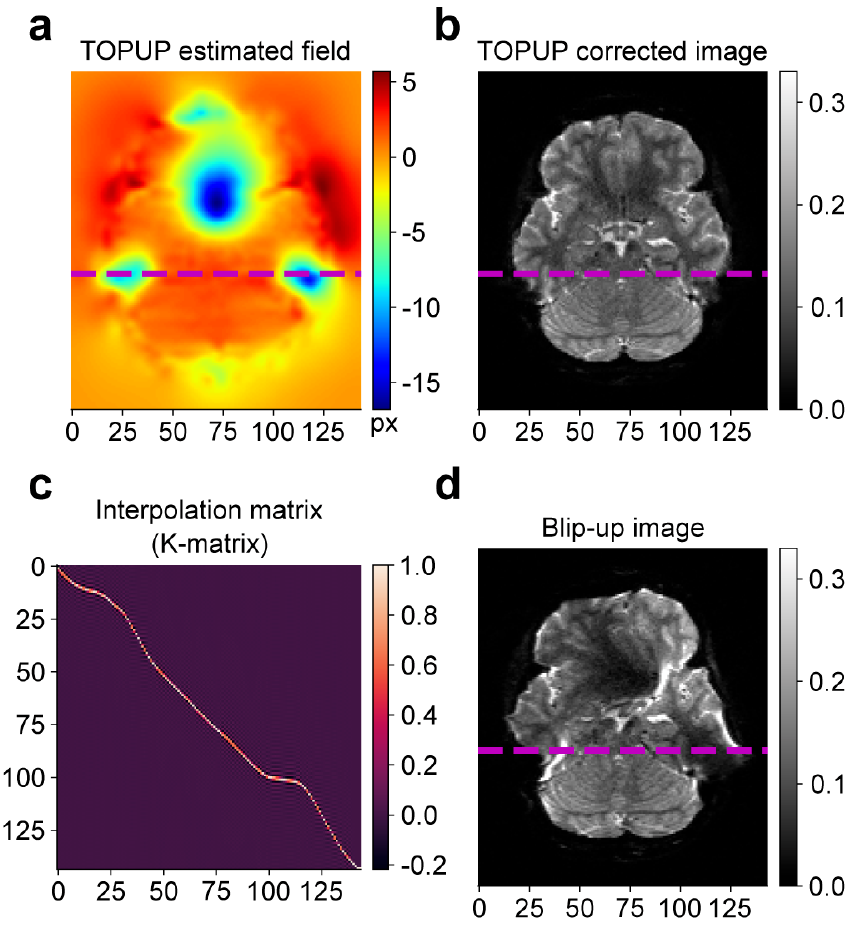}
    \caption{Illustration of the image distortion characterized by the K-matrix.
    (a) The estimated displacement field (in units of pixels) and (b) the corrected image predicted by TOPUP are shown, with the magenta dashed lines highlighting a particular row along the PE direction (RL direction).
    (c) The K-matrix formed from the field for the highlighted row and (d) the corresponding blip-up EPI image.
    The deviations of the K-matrix from the identity matrix indicate the amount and direction of distortion, as can be understood by comparing the corrected image and the blip-up image for the highlighted row.
    The labeled axes correspond to the PE direction.}
    \label{fig:kunit}
    \end{figure}

     The relationship between the anatomically-correct image and the distorted EPI image can be expressed as a linear system:
    \begin{equation}
    \label{eqn:topup_maineqn}
        \underbrace{\mathbf{\vphantom{K\mathbf{\rho}}f}}_{n_{FE} n_{PE} \times 1} = \underbrace{\mathbf{\vphantom{f\rho}K}}_{n_{FE} n_{PE} \times n_{FE} n_{PE}} \ \underbrace{\mathbf{\vphantom{fK}\rho}}_{n_{FE} n_{PE} \times 1},
    \end{equation}
    where $\mathbf{K}$ is a transformation matrix called the K-matrix, $\mathbf{\rho}$ is the vectorized anatomically-correct image, $\mathbf{f}$ is the vectorized EPI image, and $n_{PE}$ and $n_{FE}$ are the image dimensions in the PE and frequency encode (FE) directions, respectively.
    In general, $\mathbf{K}$ can be complex-valued given complex-valued images $\mathbf{\rho}$ and $\mathbf{f}$, such that it performs a phase shift as well as interpolation~\cite{Andersson_2003}.
    In practice, however, magnitude images are more commonly utilized for convenience and $\mathbf{K}$ is real-valued.
    Ignoring the distortion along the FE-direction enables block diagonalization of the K-matrix, allowing the problem to be separated across FE lines as:
    \begin{equation}
    \label{eqn:topup_k_sep}
        \underbrace{\mathbf{\vphantom{K\mathbf{\rho}}f}_{i}}_{n_{PE} \times 1} = \underbrace{\mathbf{\vphantom{f\rho}K}_{i}}_{ n_{PE} \times n_{PE}} \ \underbrace{\mathbf{\vphantom{fK}\rho}_i}_{n_{PE} \times 1},
    \end{equation}
    Here, $\mathbf{K}_{i}$, $i=1,2,\cdots,n_{FE}$, are the transformation submatrices acting along the PE-direction, and $\mathbf{\rho}_i$ and $\mathbf{f}_{i}$ are the $i^{th}$ rows of the correct image and the EPI image, respectively.
    As shown in \figureautorefname~\ref{fig:kunit}, the K-matrix describes the mapping from the correct image to the EPI image.
    Deviations of the K-matrix from the identity matrix are representative of the amount of distortion, and multiple nonzero values on the same row indicate a many-to-one mapping (i.e., pileup/dropout distortions).

    For reversed-PE acquisitions, \equationautorefname~\eqref{eqn:topup_k_sep} can be written separately for the $i^{th}$ rows of the EPI images from BU/BD acquisitions.
    In that case, the associated K-matrices $\mathbf{K}_{i,\text{BU}}$ and $\mathbf{K}_{i,\text{BD}}$ are based on the same underlying field, with the difference of utilizing the negative of the field for BD acquisition.
    
    \subsection{Classical Methods}
    
    Among classical methods for susceptibility-artifact correction in EPI, the predominant approach is correction based on reversed-PE acquisitions.
    TOPUP, a popular implementation of this approach, uses an alternating least-squares optimization to jointly solve the linear system of equations resulting from the reversed-PE acquisitions ~\cite{Smith_2004,Andersson_2003}.
    TOPUP first estimates the underlying field, which is taken as a compact linear combination of spatial basis functions across the image domain~\cite{Andersson_2003}.
    Next, transformation matrices that act on BU/BD acquisitions are generated based on the estimated field.
    Finally, to generate the anatomically correct image, unwarping is performed on BU/BD acquisitions by incorporating Jacobian modulation to compensate for intensity pileups.
    A main limitation of this method is that it relies on iterative optimization techniques that are computationally intensive.
    
    Another classical method for correcting susceptibility artifacts is B\textsubscript{0} field map based correction, which requires at least two additional acquisitions with different TE values for computing the field based on phase differences.
    This field is then used to correct the distorted EPI images by unwarping in image domain.
    However, erroneous field maps can elicit residual artifacts after correction, and phase unwrapping during field computation is prone to failure especially in regions with high B\textsubscript{0} inhomogeneities, such as air/tissue and bone/soft tissue interfaces~\cite{Zeng_2002}.
    Yet another classical method is registration-based correction, which requires an additional anatomical reference image to perform registration with the use of a cross-modal loss function~\cite{Andersson_2014_bc}.
    A distortion-free T\textsubscript{1}- or T\textsubscript{2}-weighted image typically serves as an anatomical template for the EPI image in the presence of large distortions.
    Additional constraints are often incorporated to improve solutions, including diffusion tensor~\cite{Irfanoglu_2015} and fiber orientation distributions~\cite{Qiao_2019}, alignment of cortical surfaces~\cite{Esteban_2016} and synthesized anatomical images~\cite{Li_2023}.
    Popular implementations of the aforementioned methods provided in FSL are FUGUE and FLIRT, which perform B\textsubscript{0} field map based correction and image registration based correction, respectively~\cite{Jenkinson_2001, Jenkinson_2002}.
    However, in addition to requiring auxiliary scans, these approaches fall short at capturing more intricate distortions or compensating for signal intensity variations~\cite{Tax_2022}.
    Alternatively, methods based on PSF measurements have been proposed for analytical correction based on regularized deconvolution~\cite{In_2017, Patzig_2021}, where learning-based deconvolution methods can also be adopted to improve performance~\cite{Hu_2020, Ye_2023}.
    While PSF-based methods can correct a broad range of distortions in EPI images, they require voxel-wise PSF measurements via prolonged scans that must be repeated under notable changes in k-space trajectories~\cite{Paul_2009}.
    
    \subsection{Learning-based Methods}
    
    In recent years, learning-based approaches have been adopted as a promising alternative for correction of susceptibility artifacts in EPI.
    A first group of methods have aimed to improve performance of classical methods via complementary data processing.
    Synthesis methods are applicable in cases where reversed-PE data are not available~\cite{Schilling_2019,  Schilling_2020}.
    After an undistorted EPI image is synthesized given as input a structural MR image, synthesized and acquired EPI images are processed via TOPUP to unwarp the acquired image~\cite{Schilling_2019,  Schilling_2020}.
    While suited for clinical data acquired under time limitations, synthesis methods can yield images with reduced resolution when compared to those based on reversed-PE acquisitions.
    Fiber-orientation distribution (FOD) methods use latent features of FOD images extracted from DWI data to further improve TOPUP-based correction of reversed-PE images~\cite{Qiao_2022}.
    FOD methods incorporate additional anatomical information to improve performance in problematic regions such as the brainstem.
    However, they still rely on the relatively slow TOPUP correction.
    Learning-based correction with multi-shot EPI sequences has also been considered to help minimize the distortions in acquired images.
    Low-rank reconstructions of a multi-shot EPI sequence based on simultaneous multislab acquisition has been proposed for DWI~\cite{Liao_2021}.
    Self-supervised denoising of a multi-contrast multi-shot EPI sequence based on reversed-PE acquisitions has been proposed for T\textsubscript{2}, T\textsubscript{2}\textsuperscript{*}, and susceptibility mapping~\cite{Zhang_2022}.
    Physics-driven reconstruction of an echo-shifting acquisition has been proposed for relaxometry along with B\textsubscript{0} and B\textsubscript{1} mapping~\cite{So_2022}.
    Note that these methods involve advanced pulse sequence modifications that may not be available at all sites, and  often leverage TOPUP for estimation of field maps.
    
    A second group of methods have instead aimed to improve computational efficiency over classical correction methods.
    A common framework in this domain relies on field estimation followed by unwarping of EPI images.
    Earlier studies have considered supervised methods that train network models for correction assuming availability of ground truth for undistorted EPI images~\cite{Cao_2017,Krebs_2017,Yang_2017}.
    These ground truth images are typically obtained via simulations or from classical correction methods.
    Some supervised methods further cast estimation of the displacement field from a reversed-PE image pair as an optical flow estimation problem, and later use the estimated field for correction~\cite{Dosovitskiy_2015, Ilg_2017}.
    Although supervised methods benefit from the data-driven learning capabilities of network models, reliance on the availability of undistorted EPI images limits their utility in many applications where such ground truth is not available.
    
    This has sparked interest in unsupervised methods that can learn to correct artifacts in the absence of ground truth.
    As in the case of classical methods, the predominant approach for unsupervised correction relies on reversed-PE acquisitions.
    Based on the assumption that displacements in non-PE directions are negligible~\cite{Holland_2010}, the displacement field is estimated so as to maximize the similarity of unwarped images obtained by reverse distortion on the acquired PE image pair.
    The recently proposed S-Net~\cite{Duong_2020} utilizes a 3D U-Net model~\cite{Ronneberger_2015} to predict the field, followed by unwarping using bilinear interpolation inspired by the deformable image registration method VoxelMorph~\cite{Balakrishnan_2019}.
    For unsupervised learning, S-Net uses a similarity loss taken as the local cross-correlation (LCC) between corrected BU/BD images, along with a diffusion regularizer to enforce field smoothness.
    Another recent method named Deepflow-Net~\cite{Zahneisen_2020} uses a 2D U-Net model where field estimates are produced at multiple resolutions by extracting features from various stages of the decoder~\cite{Dosovitskiy_2015, Ilg_2017}.
    Deepflow-Net performs correction via cubic interpolation and adopts a density compensation similar to TOPUP~\cite{Andersson_2003} to handle pileups.
    For unsupervised learning, Deepflow-Net uses as similarity loss the MSE between the corrected BU/BD images, along with a total variation regularizer to enforce field smoothness.
    While these seminal methods have produced promising results, they enable unsupervised learning by assessing similarity of unwarped images in opposing PE directions.
    This indirect approach omits physics-driven constraints regarding the actual EPI measurements.
    Thus, performance of the learned correction can degrade under relatively large distortions and near tissue boundaries.
    
    Here, we propose a novel unsupervised deep-learning method for artifact correction in EPI to improve performance.
    Unlike previous unsupervised methods, the proposed FD-Net method directly constrains fidelity to the actual EPI measurements.
    This constraint is introduced by integrating the forward physical model of EPI distortions observed on measured images, so FD-Net benefits from the enhanced reliability of physics-driven deep learning.
    
    \subsection{Proposed FD-Net}\label{subsec:fdnet_theory}
    
\begin{figure}[htbp]
    \centering
    \includegraphics[width=\textwidth]{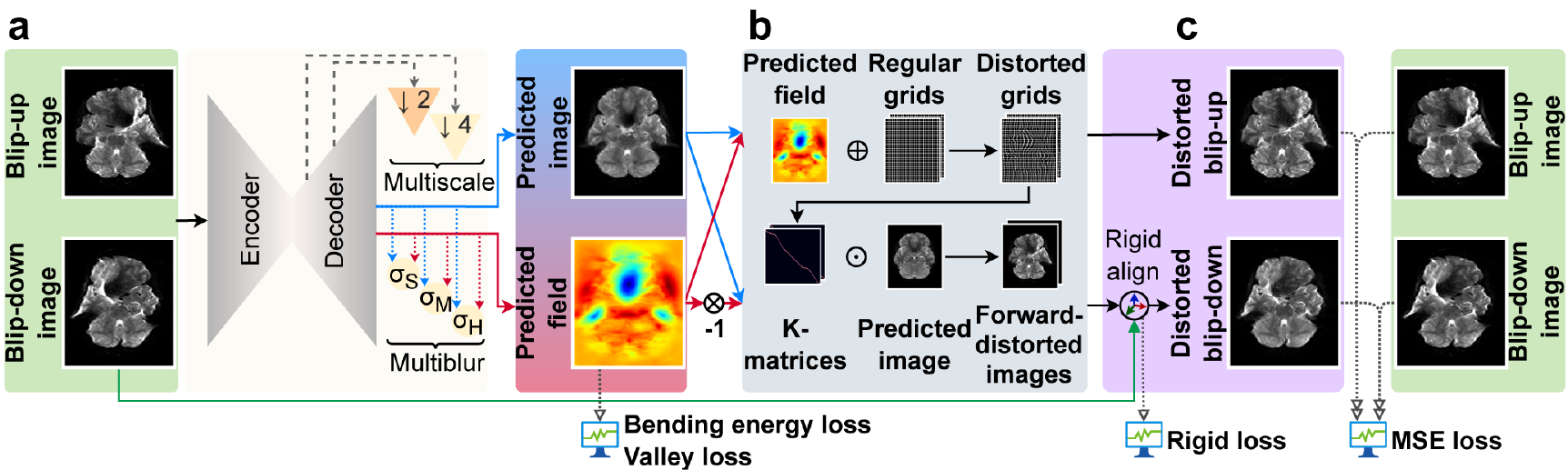}
    \caption{Overview of the proposed FD-Net.
    (a) The input distorted blip-up/blip-down images are fed through an encoder-decoder in the prediction unit, which outputs a predicted image and a predicted field with optional multiresolution (multiscale and/or multiblur) schemes.
    The field is used to formulate the bending energy loss and valley loss.
    (b) The K-Unit applies forward-distortion in each PE direction, with the field negated for one of the directions.
    (c) A rigid alignment unit is included to improve registration, with the rigid loss formulated from the transformation parameters.
    The forward-distorted images are compared with the input images (redisplayed here for convenience) to formulate the MSE loss.
    Training is performed over the aggregate of the shown losses.}
    \label{fig:overview}
    \end{figure}

    FD-Net is a novel unsupervised forward-distortion model that explicitly enforces measurement fidelity for enhanced correction performance, as outlined in \figureautorefname~\ref{fig:overview}.
    The prediction unit, shown in \figureautorefname~\ref{fig:overview}a, uses a 2D U-Net to produce both a predicted field and a predicted anatomically-correct image from the input reversed-PE images.
    In contrast to unwarping-based methods that produce separate unwarped images for BU/BD acquisitions, predicting a single correct image can offer SNR benefits analogously to the sensitivity-encoding approaches in parallel imaging~\cite{Pruessmann_1999}.

    The K-Unit in FD-Net, illustrated in \figureautorefname~\ref{fig:overview}b, forward-distorts the predicted anatomically-correct image using the predicted field to reconstruct the input PE images.
    The BU acquisition is reconstructed using the estimated field, whereas the BD acquisition is reconstructed using the negative of the estimated field.
    Distortions are efficiently emulated using the K-Unit that embodies a simple matrix multiplication with a separable formulation as in \equationautorefname~\eqref{eqn:topup_k_sep}.
    Afterwards, fidelity between reconstructed and measured data is enforced using a multiresolution scheme.
    
    The rigid alignment unit in \figureautorefname~\ref{fig:overview}c allows compensation for small movements between the input PE image in one direction (BD acquisition in this case) and its corresponding forward-distorted image.
    This allows the network to focus on displacements that are due to off-resonance via the field-based formulation of the K-Unit.
    
        \subsubsection{Forward-Distortion with K-Unit}\label{subsubsec:forward_distortion_with_kunit}
        
\begin{figure}[htbp]
        \centering
        \includegraphics[width=0.5\textwidth]{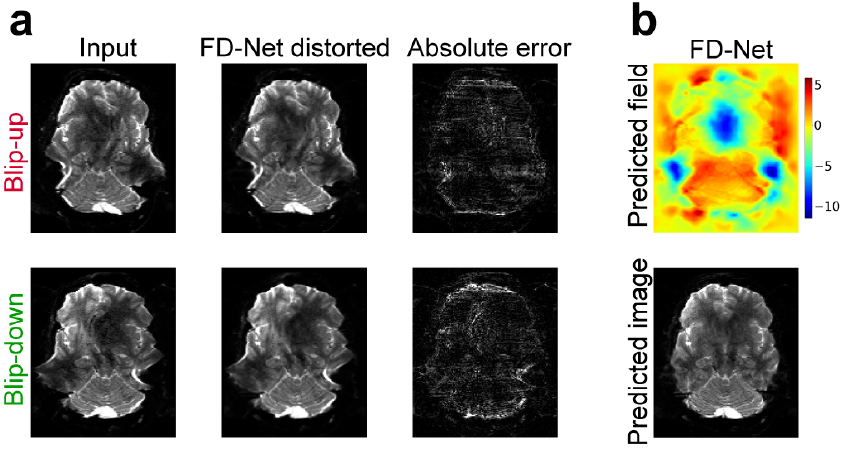}
        \caption{Example of forward-distortion by using the K-Unit in FD-Net.
        (a) The input blip-up and blip-down EPI images are compared with the forward-distortion results of FD-Net.
        The intensities in absolute error maps are scaled up $2.5\times$ for improved visualization.
        (b) The predicted field and predicted image outputs from FD-Net, which are input to the K-Unit to obtain the forward-distorted images in (a).}
        \label{fig:kunit_fdnet_ex}
        \end{figure}

        The K-Unit in FD-Net performs forward-distortion on the estimated anatomically-correct image using the estimated field, as illustrated in \figureautorefname~\ref{fig:kunit_fdnet_ex}.
        The steps described below are given for the BU direction for brevity, but they are similarly conducted for the BD direction, with the difference of utilizing the negative of the displacement field.
        First, a uniform spatial grid $\mathbf{X}_{\text{grid}}$ is formed:
        \begin{equation}
        \label{eqn:xgrid}
            \underbrace{\mathbf{X}_{\text{grid}}}_{n_{PE} \times n_{PE}} =
            \begin{bmatrix}
                1 & \cdots & 1\\
                2 & \cdots & 2\\
                \vdots &  & \vdots\\
                n_{PE} & \cdots & n_{PE}
            \end{bmatrix} .
        \end{equation}
        The distorted grid after interpolation, $\mathbf{X}_{i,\text{BU}}$, is formed by determining the new grid location for each pixel from the shift amount given in the displacement field, i.e.,
        \begin{equation}
        \label{eqn:xinterp}
            \underbrace{\mathbf{X}_{i,\text{BU}}}_{n_{PE} \times n_{PE}} =
            \begin{bmatrix}
                \mathbf{O}_{\text{field}}(i,1) + 1 & \cdots & \mathbf{O}_{\text{field}}(i,n_{PE}) + n_{PE}\\
                \mathbf{O}_{\text{field}}(i,1) + 1 & \cdots & \mathbf{O}_{\text{field}}(i,n_{PE}) + n_{PE}\\
                \vdots & & \vdots\\
                \mathbf{O}_{\text{field}}(i,1) + 1 & \cdots & \mathbf{O}_{\text{field}}(i,n_{PE}) + n_{PE}
            \end{bmatrix},
        \end{equation}
        where $\mathbf{O}_{\text{field}}$ is the estimated field output of FD-Net in units of pixels and $i=1,2,\ldots,n_{FE}$ is the row index over the FE direction.
        For practical purposes, each entry in $\mathbf{X}_{i,\text{BU}}$ is kept limited between $1$ and $n_{PE}$ (i.e., clipped to the valid range of interpolation).
        Taking the difference between the two grids and then applying an interpolation kernel, $\kappa(\xi)$, gives us the K-matrix that will act on the $i^{th}$ row as follows:
        \begin{equation}
        \label{eqn:k_xinterp_xgrid}
            \underbrace{\mathbf{K}_{i,\text{BU}}}_{n_{PE} \times n_{PE}} = \kappa\left(\mathbf{X}_{i,\text{BU}} - \mathbf{X}_{grid} \right) .
        \end{equation}
        Using this K-matrix, the $i^{th}$ row of the forward-distorted image is reconstructed via a matrix multiplication:
        \begin{equation}
        \label{eqn:odist_oimage}
            \underbrace{\left[\mathbf{O}_{\text{dist,BU}}^{T}\right]_{i}}_{n_{PE} \times 1} =  \mathbf{K}_{i,\text{BU}} \ \underbrace{\left[\mathbf{O}_{\text{image}}^{T}\right]_{i}}_{n_{PE} \times 1},
        \end{equation}
        where $(\cdot)^T$ denotes matrix transpose, $[\cdot]_i$ denotes the $i^{th}$ column of a matrix, and $\mathbf{O}_{\text{image}}$ is the predicted anatomically-correct image.
        Finally, the forward-distorted image $\mathbf{O}_{\text{dist,BU}}$ can be formed by stacking the individually distorted rows:
        \begin{equation}
        \label{eqn:odist_stack}
            \underbrace{\mathbf{O}_{\text{dist,BU}}}_{n_{FE} \times n_{PE}} =
            \left[\begin{array}{c|c|c|c}
                \left[\mathbf{O}_{\text{dist,BU}}^{T}\right]_{1} &
                \left[\mathbf{O}_{\text{dist,BU}}^{T}\right]_{2} &
                \cdots &
                \left[\mathbf{O}_{\text{dist,BU}}^{T}\right]_{n_{FE}}
            \end{array}\right]^{T}.
        \end{equation}
        Note that multiplication with K-matrix rows performs an interpolation across pixel neighborhoods with intensity modulations, so it can emulate signal pileups/dropouts.

        \subsubsection{Network Architecture}\label{subsubsec:network_architecture_thoery}
        
\begin{figure}[htbp]
        \centering
        \includegraphics[width=\textwidth]{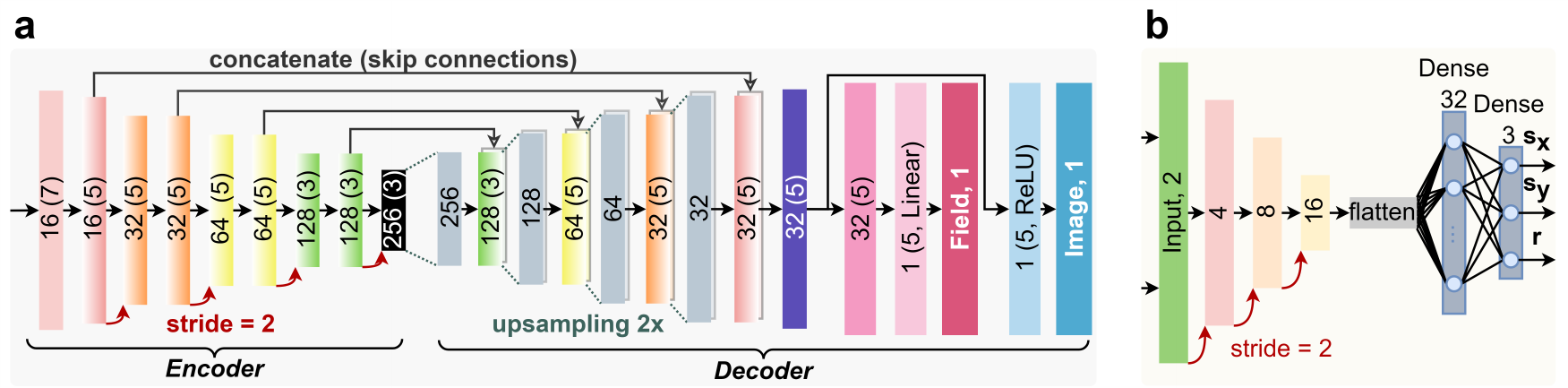}
        \caption{Details of the network architecture of FD-Net.
        (a) The encoder-decoder in the prediction unit is outlined with blocks representing the convolutional steps. Convolution with stride 2 is used to reduce dimensionality in the encoder steps, while upsampling by 2 is used to increase it in the decoder steps. Skip connections are introduced to facilitate information flow and improve gradient propagation by concatenating the encoder representations to the corresponding stages in the decoder. The numbers inside the boxes denote feature dimensions, with the numbers in brackets indicating filter kernel sizes. Leaky ReLU activation with slope coefficient of 0.2 is used, unless otherwise indicated.
        (b) A rigid alignment unit is used to align one of the forward-distorted images (blip-down distorted image in this case) to its corresponding input EPI image. The first stage encodes the images using convolutional layers with stride of 2, shown as boxes with the numbers inside indicating the feature dimensions. The output of the convolutional stage is flattened and passed through a dense layer with 32 neurons and another with 3 neurons. The final output comprises the 3 parameters required for the rigid transformation to be applied to the forward-distorted image.}
        \label{fig:networks}
        \end{figure}

        The architecture of FD-Net is detailed in \figureautorefname~\ref{fig:networks}.
        As depicted in \figureautorefname~\ref{fig:networks}a, the encoder in the prediction unit projects input reversed-PE images onto a latent representation across multiple stages.
        The receptive field is progressively refined by decreasing kernel size and using convolution with stride 2 for downsampling.
        The decoder then resolves the predicted field and predicted image from the latent representation through multiple stages of convolutional filtering and upsampling.
        Feature maps from the encoder stages are projected onto the decoder through skip connections to improve information flow.

        A rigid-body motion may occur between the BU and BD acquisitions.
        As shown in \figureautorefname~\ref{fig:networks}b, the rigid alignment unit in FD-Net applies motion-related transformations on one of the forward-distorted images only (BD distorted image in this case).
        This unit receives as input the measured BD acquisition along with the respective forward-distorted image, and uses convolutional and densely connected layers to predict the motion parameters $s_x$, $s_y$, and $r$, which capture the x-axis shift, y-axis shift, and in-plane rotation, respectively.
        These parameters are then used to apply a rigid transformation to the BD distorted image to improve its alignment with the corresponding BD acquisition.
        Note that a similar rigid alignment is also performed in TOPUP, and it offloads some burden from the non-rigid field-based alignment by accounting for subject movement between the two reversed-PE acquisitions.

\begin{figure}[htbp]
        \centering
        \includegraphics[width=\textwidth]{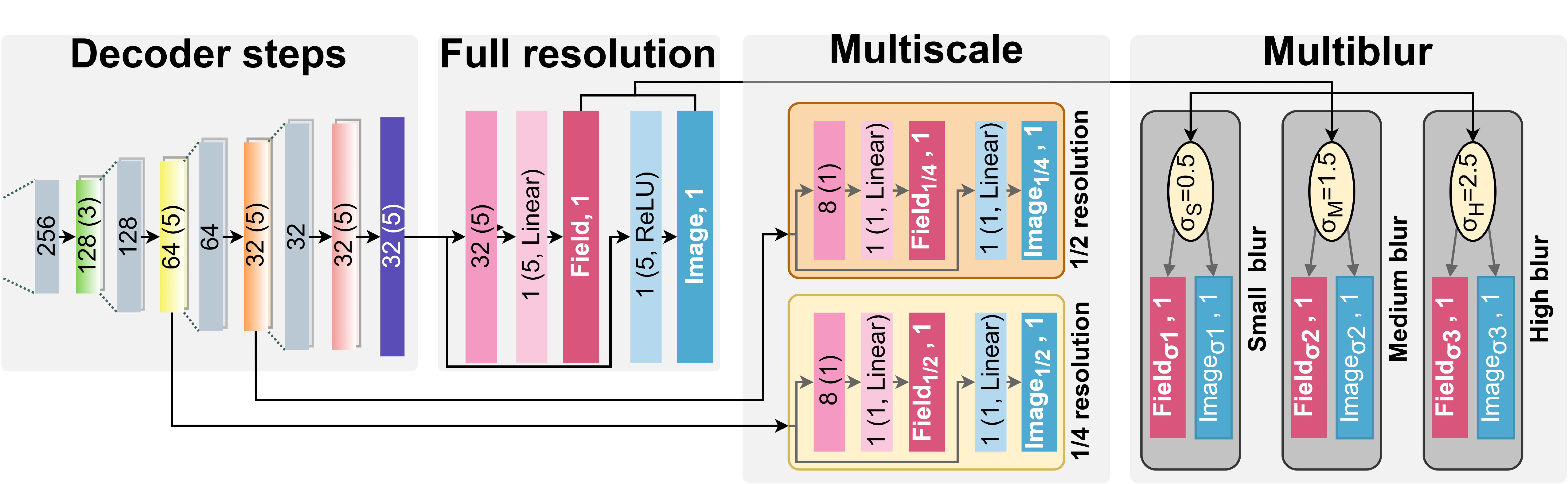}
        \caption{Illustration of two different multiresolution approaches, multiscale and multiblur, considered for FD-Net.
        The last stages of the decoder that generate the full resolution image and field are also shown for clarity.
        The multiscale approach relies on forming an output image and field at a lower dimensional scale, using appropriate convolutional steps to produce the outputs at 1/2 and 1/4 scale in this case.
        The numbers inside the boxes denote feature dimensions, with the numbers in brackets indicating filter kernel sizes.
        Leaky ReLU activation with slope coefficient of 0.2 is used, unless otherwise indicated.
        For the multiblur case, Gaussian blur kernels are applied to the full resolution outputs to create increasingly blurred results.
        Small, medium, and high blur amounts of $\sigma_{S}=0.5$, $\sigma_{M}=1.5$, and $\sigma_{H}=2.5$ are used, respectively, with a Gaussian kernel size of $\lceil 4\sigma_m\rceil$ in each case.
        The results of all the incorporated multiresolution levels are then passed through the K-Unit shown in \figureautorefname~\ref{fig:kunit}, contributing to the overall loss in a regularizing manner.}
        \label{fig:multires}
        \end{figure}

        As illustrated in \figureautorefname~\ref{fig:multires}, FD-Net adopts a multiresolution scheme to improve performance by enforcing consistency across different spatial resolutions, in principle leading to faster convergence and more reliable performance.
        In FD-Net, we refer to the evaluation at different spatial scales as multiscale and at different spatial blurs as multiblur.
        For multiscale, field and anatomically-correct image estimates are produced at multiple spatial resolutions by extracting outputs from different stages of the decoder.
        For multiblur, the full resolution outputs are blurred with Gaussian kernels at varying standard deviations.
        In both cases, the estimates obtained at multiple scales/blurs are processed with the K-Unit after proper scaling of their contribution to the overall loss function.

        \subsubsection{Network Loss}\label{subsubsec:network_loss_theory}
        
        The overall loss function for FD-Net is given as:
        \begin{equation}
        \label{eqn:fdnet_maineqn}
            \mathcal{L}_{FD-Net} = \sum_{m} \omega_m \left[ \mathcal{L}_{MSE}^{(m)} + \lambda_m \left( \mathcal{L}_{BE}^{(m)} + 10^{3}\mathcal{L}_{valley}^{(m)} \right) \right] \ + \gamma\mathcal{L}_{rigid},
        \end{equation}
        where $m$ is the index of multiresolution step, $\omega_m$ and $\lambda_m$ are the weighting and regularization parameter over the smoothness of the field for step $m$, superscript $(m)$ denotes the version of a parameter at step $m$, and $\gamma$ is the weight of the rigid alignment loss.
        Here, the first term denotes the sum of reconstruction losses, while the second term denotes the rigid loss, described in detail below.
        
        First, $\mathcal{L}_{MSE}^{(m)}$ is MSE between the measured and forward-distorted images averaged across the two PE directions at the $m^{\text{th}}$ step:
        \begin{equation}
        \label{eqn:fdnet_mse}
           \mathcal{L}_{MSE}^{(m)} = \frac{1}{2 n_{PE}^{(m)} n_{FE}^{(m)}} \left[\sum_{\mathbf{p} \in \Omega} \left( \mathbf{O}_{\text{dist,BU}}^{(m)}(\mathbf{p}) - \mathbf{I}_{\text{im,BU}}^{(m)}(\mathbf{p}) \right)^2  + \sum_{\mathbf{p} \in \Omega} \left( \mathbf{O}_{\text{dist,BD}}^{(m)}(\mathbf{p}) - \mathbf{I}_{\text{im,BD}}^{(m)}(\mathbf{p}) \right)^2  \right],
        \end{equation}
        where $\mathbf{I}_{\text{im,BU}}$ and $\mathbf{I}_{\text{im,BD}}$ are the input EPI images for BU and BD acquisitions, respectively.
        For the multiscale scheme, these images are downsampled properly to avoid aliasing artifacts.
        
        Next, $\mathcal{L}_{BE}^{(m)}$ is the bending energy regularizer~\cite{Staring_2007} over the field at each step $m$ expressed as:
        \begin{equation}
        \label{eqn:fdnet_be}
            \mathcal{L}_{BE}^{(m)} = \sum_{\mathbf{p} \in \Omega}
            \left( \frac{\partial^2}{\partial x^2} \mathbf{O}_{\text{field}}^{(m)}(\mathbf{p}) \right)^2 + 
            \left( \frac{\partial^2}{\partial y^2} \mathbf{O}_{\text{field}}^{(m)}(\mathbf{p}) \right)^2 + 
            \left( \frac{\partial^2}{\partial xy} \mathbf{O}_{\text{field}}^{(m)}(\mathbf{p}) \right)^2 + 
            \left( \frac{\partial^2}{\partial yx} \mathbf{O}_{\text{field}}^{(m)}(\mathbf{p}) \right)^2 .
        \end{equation}
        In practice, first- and second-order finite differences are used to approximate the gradients~\cite{Fornberg_1988}.
        
        $\mathcal{L}_{valley}^{(m)}$ is the valley loss for the field to prevent the overall loss function from exploding in earlier training iterations~\cite{Zahneisen_2020}, and is given as:
        \begin{equation}
       \label{eqn:fdnet_valley}
           \mathcal{L}_{valley}^{(m)} = \sum_{\mathbf{p} \in \Omega} \max \left( \left\vert \mathbf{O}_{\text{field}}^{(m)}(\mathbf{p}) \right\vert - \tau_m, 0 \right),
        \end{equation}
        where $\tau_m$ is a chosen threshold of maximum permissible field swing in units of pixels.
        $\mathcal{L}_{valley}^{(m)}$ sums the excess amount of field swing values when their magnitudes exceed $\tau_m$.
        These cases are penalized heavily by weighting $\mathcal{L}_{valley}^{(m)}$ with a large constant in \equationautorefname~\eqref{eqn:fdnet_maineqn}.
        In later stages of training, the effect of $\mathcal{L}_{valley}^{(m)}$ is negligible once the network converges towards reasonable solutions.
    
        Finally, $\mathcal{L}_{rigid}$ is the rigid loss to find the smallest possible rigid transformation parameters for the alignment of measured and forward-distorted BD images, and is defined as follows:
        \begin{equation}
        \label{eqn:fdnet_rigid}
            \mathcal{L}_{rigid} = s_x^2 + s_y^2 + r^2 .
        \end{equation}
        Because the same rigid alignment applies to all multiresolution steps, a single rigid loss term is included in \equationautorefname~\eqref{eqn:fdnet_maineqn}.
\section{Methods}\label{sec:methods}

    \subsection{Experimental Dataset and Setup}\label{subsec:experimental_dataset}

    For the experiments in this work, unprocessed DWI data from HCP 1200 Subjects Data Release were used~\cite{VanEssen_2013}.
    The images were acquired on a 3T MRI scanner (Siemens Skyra ``Connectom''), using a multiband diffusion sequence with ss-EPI readouts in right-to-left (RL) and left-to-right (LR) reversed-PE polarities~\cite{HCP_AppendixI_2017}.
    Other imaging parameters included: {$210 \times 180$}~{mm\textsuperscript{2}} FOV, $1.25$~mm isotropic resolution, averages~{$= 1$}, multi-band acceleration factor~$3$; TR/TE~{$= 5520/89.50$ ms}, flip angle~{$= 78^{\circ}$}; {$168 \times 144$}~acquisition matrix, bandwidth~{$= 1488$ Hz/Px}, EPI factor~{$= 144$}, echo spacing~{$= 0.78$ ms}, and $6/8$~phase partial Fourier acquisition.
    
    A total of 24 subjects were selected randomly from the HCP database, with 12 reserved for training, 4 for validation, and 8 for testing.
    For each subject, a single b0-volume consisting of 111 axial slices with 168$\times$144 image matrix was utilized.
    To obtain reference corrected images, the TOPUP method was applied on the data following the recommended guidelines by the toolbox.
    
    All networks were implemented in Keras with Tensorflow backend, on a machine with NVIDIA RTX 3070 GPU.
    Training was performed with the Adam optimizer for a learning rate of $10^{-4}$ and a maximum of 1000 epochs, with early stopping when the change in the validation loss between consecutive epochs in the validation set fell below a threshold of $10^{-6}$.

    \subsection{FD-Net Implementation}\label{subsec:proposed_method}

    The columns of the K-matrix in the K-Unit were generated using a sinc kernel, i.e., $\kappa(\xi) = \text{sinc}(\xi)$.
    All convolutional layers in the encoder-decoder (i.e., U-Net) utilized Leaky Rectified Linear Unit (ReLU) activation with a slope coefficient $\alpha=0.2$, except at the final steps of the decoder as indicated in \figureautorefname~\ref{fig:networks}a; the predicted image was output via a convolutional layer with ReLU activation and the predicted field was output via a convolutional layer with linear activation.
    For the multiscale case, convolutional layers akin to the full resolution case were employed to form the predicted field and image at 1/2 and 1/4 of the full scale.
    For the multiblur case, the full resolution output was blurred with Gaussian kernels of standard deviation $\sigma_{m}$ and width $\lceil 4\sigma_{m} \rceil$.
    Three different blur levels were used: small (S), medium (M), and high (H) blurs of $\sigma_{S}=0.5$, $\sigma_{M}=1.5$, and $\sigma_{H}=2.5$, respectively.
    
    \subsection{Competing Methods}\label{subsec:competing_methods}
    
    Two unsupervised learning-based methods, S-Net and Deepflow-Net, were implemented for comparison.
    In addition, a supervised method was implemented to serve as a baseline for FD-Net.
    Implementations of competing methods were maintained as consistent to FD-Net as possible to facilitate fair comparisons:
    
    \begin{enumerate}
        \item \textit{S-Net:} S-Net was implemented using a 2D U-Net.
        Only the field head at the end of the decoder in \figureautorefname~\ref{fig:networks}a was necessary and correction was performed using a modified K-Unit approach as follows:
        \begin{equation}
        \label{eqn:ounwarp_ipe}
                \underbrace{\left[\mathbf{O}_{\text{unwarp,BU}}^{T}\right]_{i}}_{n_{PE} \times 1} = 
                \mathbf{K}_{i,\text{BU}}^{T} \ \underbrace{\left[\mathbf{I}_{\text{im,BU}}^{T}\right]_{i}}_{n_{PE} \times 1},
        \end{equation}
        Here, $\mathbf{O}_{\text{unwarp,BU}}$ denotes the unwarped BU image.
        Note that no density compensation was incorporated by Duong~et~al.~\cite{Duong_2020}.
        Similarly, by transposing $\mathbf{K}_{i,\text{BU}}$, a standard unwarping interpolation was performed without density compensation.
        The BD acquisition was similarly treated, with the K-matrix formed after negation of the field.
        The average of the unwarped BU/BD images was taken as the corrected image.
        For training, LCC of the unwarped BU/BD images was utilized for similarity loss~\cite{Duong_2020,Balakrishnan_2019}.
        In place of the diffusion regularizer in~\cite{Duong_2020}, bending energy from \equationautorefname~\eqref{eqn:fdnet_be} was used to facilitate comparison with FD-Net.
        In addition, the rigid alignment unit was utilized and the rigid loss from \equationautorefname~\eqref{eqn:fdnet_rigid} was incorporated.
        
        \item \textit{Deepflow-Net:} Deepflow-Net was implemented using a 2D U-Net.
        Only the field head at the end of the decoder in \figureautorefname~\ref{fig:networks}a was needed and density-compensated correction was performed based on a modified K-Unit approach.
        The K-matrix for the BU acquisition was multiplied with an image of 1's, $\mathbf{1}$, to produce a density pileup map $\mathbf{W}_{\text{BU}}$.
        This map was inverted and used to weight the input PE image to enable density compensation akin to Zahneisen~et~al.~\cite{Zahneisen_2020}:
        \begin{equation}
        \label{eqn:ounwarp_ipe_w}
            \underbrace{\left[\mathbf{O}_{\text{unwarp,BU}}^{T}\right]_{i}}_{n_{PE} \times 1} = 
            \mathbf{K}_{i,\text{BU}}^{T} \ \underbrace{\left( \left(\mathbf{1} \oslash \mathbf{W}_{\text{BU}}\right) \odot \left[\mathbf{I}_{\text{im,BU}}^{T}\right]_{i} \right)}_{n_{PE} \times 1} ,
        \end{equation}
        where
        \begin{equation}
            \underbrace{\mathbf{W}_{\text{BU}}}_{n_{PE} \times 1} = \mathbf{K}_{i,\text{BU}} \underbrace{\mathbf{1}}_{n_{PE} \times 1} .
        \end{equation}
        Here, $\oslash$ and $\odot$ denote Hadamard division and product, respectively, and $\left(\mathbf{1} \oslash \mathbf{W}_{\text{BU}}\right)$ is limited in $[0,1]$ to decrease the intensity in pileup regions~\cite{Zahneisen_2020}.
        The same process was also followed for the BD acquisition, with the K-matrix formed after negation of the field.
        In contrast to \equationautorefname~\eqref{eqn:ounwarp_ipe}, \equationautorefname~\eqref{eqn:ounwarp_ipe_w} applies density compensation together with unwarping.
        The average of the two unwarped images was used as the corrected image.
        The same multiscale strategy as in FD-Net was adopted.
        MSE between the unwarped BU/BD images was used as the similarity loss.
        In place of TV regularization in Zahneisen~et~al.~\cite{Zahneisen_2020}, bending energy loss was applied for the field as in  \equationautorefname~\eqref{eqn:fdnet_be}.
        The rigid alignment unit was also incorporated along with its loss term.
    
        \item \textit{Supervised Baseline:} Finally, a supervised baseline was trained with an architecture identical to that of FD-Net, with the exception of the loss being fully supervised.
        For this purpose, MSE between the network predicted field/image and the results from TOPUP was utilized.
    \end{enumerate}

    \subsection{Quantitative Assessments}\label{subsec:quantitative_assessments}
    
    The qualities of the predicted image and field were assessed via Peak SNR (PSNR) and Structural Similarity Index Measure (SSIM) metrics, with the TOPUP results taken as reference.
    Before computing PSNR and SSIM, the field generated by each method was masked via a median Otsu threshold over the TOPUP image to remove background regions from consideration~\cite{Garyfallidis_2014}.
    
    For all methods, hyperparameters were chosen empirically to maximize PSNR and SSIM over the 4 subjects reserved as validation data.
    The selected hyperparameters are provided in \tableautorefname~\ref{tab:params_comp}.
    Performance assessments were reported on independent test data.

    \begin{table}[htbp]
    \caption{Hyperparameter choices for the proposed FD-Net and the competing methods.
    For each method, irrelevant hyperparameters are marked with a dash ($-$).
    The hyperparameters considered are: $\lambda$ for field smoothness regularization, $\omega$ for multiresolution weighting parameter, $\gamma$ for rigid loss, and $\tau$ for valley loss threshold.
    $\omega$ is split into its constituent full resolution (``FR''), multiscale (1\textfractionsolidus 2 and 1\textfractionsolidus 4 scale), and multiblur (S, M, and H) components.}
    \centering
    \begin{tabular}{ *{3}{l} *{8}{c}}
        \toprule
        
        \multicolumn{1}{c}{Methods}
        &\multicolumn{1}{c}{}
        &\multicolumn{1}{c}{$\lambda$}
        &\multicolumn{6}{c}{$\omega$}
        &\multicolumn{1}{c}{$\gamma$}
        &\multicolumn{1}{c}{$\tau$}\\
        
        \cmidrule{4-9}
    
        \multicolumn{3}{c}{}
        &$FR$
        &1\textfractionsolidus 2
        &1\textfractionsolidus 4
        &S
        &M
        &H
        &\multicolumn{1}{c}{}
        &\multicolumn{1}{c}{}\\
        
        \midrule
    
        Proposed FD-Net
        &\multicolumn{1}{c}{}
        &$10^{-5}$
        &$0.4$
        &$-$
        &$-$
        &$0.3$
        &$0.2$
        &$0.1$
        &$0.01$
        &$32$\\
        
        Deepflow-Net
        &\multicolumn{1}{c}{}
        &$10^{-5}$
        &$0.6$
        &$0.3$
        &$0.1$
        &$-$
        &$-$
        &$-$
        &$0.01$
        &$32$\\
        
        S-Net
        &\multicolumn{1}{c}{}
        &$10$
        &$1.0$
        &$-$
        &$-$
        &$-$
        &$-$
        &$-$
        &$0.01$
        &$-$\\
        
        Supervised baseline
        &\multicolumn{1}{c}{}
        &$-$
        &$1.0$
        &$-$
        &$-$
        &$-$
        &$-$
        &$-$
        &$-$
        &$-$\\
        
        \bottomrule
    \end{tabular}
    \label{tab:params_comp}
    \end{table}

\section{Results}\label{sec:results}

    \subsection{Computation Time}\label{subsec:computation_time}
    
    All competing methods provided substantial computational advantage over TOPUP.
    Correction of a volume took on average ${\sim}7.5$~sec for each network considered.
    In contrast, TOPUP took on average ${\sim}3086$~sec (${\sim}51.5$~min) to predict the field and an additional ${\sim}6$~sec to apply correction.
    Thus, network-based artifact correction enabled significant speed up over classical methods.

    \subsection{Ablation Studies for FD-Net}\label{subsec:ablation_fd-net}

    The choice of multiresolution strategy for FD-Net was first considered, followed by an ablation study on the combination of multiresolution components.
    The parameters were chosen empirically, with the purpose of maximizing quantitative image quality metrics with respect to TOPUP over the predicted field/image.
    Lastly, an ablation study was conducted to evaluate the contribution of each loss term in \equationautorefname~\eqref{eqn:fdnet_maineqn}.
    
        \subsubsection{Multiresolution Ablation Study for FD-Net}\label{subsec:multiresolution_ablation_study}

        FD-Net was trained and subsequently evaluated for each multiresolution strategy, alongside a strategy with no multiresolution.
        The performances of the multiscale and multiblur schemes, as well their combination, were compared to determine the best multiresolution strategy.
        The hyperparameters chosen for each multiresolution scheme considered are provided in \tableautorefname~\ref{tab:params_multires}.
        PSNR and SSIM metrics are listed in \tableautorefname~\ref{tab:results_multires}.
        Overall, introducing a multiblur strategy provides a performance boost.
        Using the multiblur strategy, the image quality is improved by $0.67\text{dB}$ PSNR/$2.11\%$ SSIM, and the field quality is improved by $1.68\text{dB}$ PSNR/$2.94\%$ SSIM over the no multiresolution case.
        In contrast, the multiscale strategy underperforms in comparison to both the multiblur and the no multiresolution cases.
        A combination of multiblur and multiscale strategies does not improve over the multiblur case either, indicating that multiblur alone is sufficient to boost performance.
        Hence, the multiblur strategy was selected for FD-Net.

        \begin{table}[htbp]
        \caption{Hyperparameter choices for the multiresolution strategy ablation study for FD-Net.
        For each multiresolution strategy, irrelevant hyperparameters are marked with a dash ($-$).
        The multiresolution weighting parameter, $\omega$, is split into its constituent weights for full resolution (``FR''), multiscale (1\textfractionsolidus 2 and 1\textfractionsolidus 4 scale), and multiblur (S, M, and H) components.
        The other hyperparameters are kept fixed: $\lambda$ for field smoothness regularization, $\gamma$ for rigid loss, and $\tau$ for valley loss threshold.}
        \centering
        \begin{tabular}{ *{3}{l} *{8}{c}}
            \toprule
            
            \multicolumn{1}{c}{Multiresolution scheme}
            &\multicolumn{1}{c}{}
            &\multicolumn{1}{c}{$\lambda$}
            &\multicolumn{6}{c}{$\omega$}
            &\multicolumn{1}{c}{$\gamma$}
            &\multicolumn{1}{c}{$\tau$}
            \\
            
            \cmidrule{4-9}
        
            \multicolumn{3}{c}{}
            &$FR$
            &1\textfractionsolidus 2
            &1\textfractionsolidus 4
            &S
            &M
            &H
            &\multicolumn{1}{c}{}
            &\multicolumn{1}{c}{}
            \\
            
            \midrule
        
            No multiresolution
            &\multicolumn{1}{c}{}
            &$10^{-5}$
            &$1.0\phantom{0}$
            &$-$
            &$-$
            &$-$
            &$-$
            &$-$
            &$0.01$
            &$32$
            \\
            
            Multiscale
            &\multicolumn{1}{c}{}
            &
            &$0.6\phantom{0}$
            &$0.3\phantom{0}$
            &$0.1\phantom{0}$
            &$-$
            &$-$
            &$-$
            &
            &
            \\
            
            Multiblur
            &\multicolumn{1}{c}{}
            &
            &$0.4\phantom{0}$
            &$-$
            &$-$
            &$0.3\phantom{0}$
            &$0.2\phantom{0}$
            &$0.1\phantom{0}$
            &
            &
            \\
            
            Multiscale\,\&\,multiblur
            &\multicolumn{1}{c}{}
            &
            &$0.5\phantom{0}$
            &$0.15$
            &$0.05$
            &$0.15$
            &$0.1\phantom{0}$
            &$0.05$
            &
            &
            \\
            
            \bottomrule
        \end{tabular}
        \label{tab:params_multires}
        \end{table}

        \begin{table}[htbp]
        \caption{Performance comparison of multiresolution strategies in FD-Net.
        PSNR and SSIM metrics are reported as mean (SD) across subjects. 
        Bold font denotes the best performing strategy.
        The multiblur strategy is chosen.}
        \centering
        \begin{tabular}{ *{1}{l} *{6}{c}}
            \toprule
            
            \multicolumn{1}{c}{Multiresolution scheme}
            &\multicolumn{1}{c}{}
            &\multicolumn{2}{c}{Image quality}
            &\multicolumn{1}{c}{}
            &\multicolumn{2}{c}{Field quality}\\
            
            \cmidrule{3-4} \cmidrule{6-7}
        
            \multicolumn{2}{l}{}
            &PSNR [dB]
            &SSIM [\%]
            &\multicolumn{1}{c}{}
            &PSNR [dB]
            &SSIM [\%]\\
            
            \midrule
        
            No multiresolution
            &\multicolumn{1}{c}{}
            &$30.62\ (2.82)$
            &$84.60\ (11.72)$
            &\multicolumn{1}{c}{}
            &$20.80\ (6.11)$
            &$80.15\ (11.85)$\\

            Multiscale
            &\multicolumn{1}{c}{}
            &$30.31\ (2.67)$
            &$83.91\ (11.72)$
            &\multicolumn{1}{c}{}
            &$20.00\ (6.14)$
            &$79.24\ (12.44)$\\

            \textbf{Multiblur}
            &\multicolumn{1}{c}{}
            &$\mathbf{31.29\ (2.79)}$
            &$\mathbf{86.71\ (11.58)}$
            &\multicolumn{1}{c}{}
            &$\mathbf{22.48\ (5.69)}$
            &$\mathbf{83.09\ (10.37)}$\\
            
            Multiscale\,\&\,multiblur
            &\multicolumn{1}{c}{}
            &$30.88\ (2.78)$
            &$85.84\ (11.79)$
            &\multicolumn{1}{c}{}
            &$21.67\ (5.82)$
            &$82.52\ (10.42)$\\
            
            \bottomrule
        \end{tabular}
        \label{tab:results_multires}
        \end{table}
    
        Next, combinations of three different blur amounts were considered for the multiblur case, with hyperparameters as listed in \tableautorefname~\ref{tab:params_multiblur}.
        The results in \tableautorefname~\ref{tab:results_multiblur} show that including two or more blur stages boosts performance.
        Incorporating all blur stages (i.e., S-M-H multiblur combination) provides the best results, improving the image quality by $2.28\text{dB}$ PSNR/$5.23\%$ SSIM and the field quality by $5.99\text{dB}$ PSNR/$8.80\%$ SSIM over the worst performing M multiblur scheme.
        Hence, the S-M-H multiblur combination was chosen for FD-Net, as it provides a reliable generalization by incorporating all blur stages.

        \begin{table}[htbp]
        \caption{Hyperparameter choices for the multiblur scheme ablation study for FD-Net.
        Combinations over three different blur amounts are considered: small (S), medium (M), and high (H) blurs of $\sigma_{S}=0.5$, $\sigma_{M}=1.5$, and $\sigma_{H}=2.5$, respectively. 
        For each combination, excluded blurs are marked with a dash ($-$). The multiresolution weighting parameter, $\omega$, is split into its constituent weights for full resolution (``FR'') and multiblur (S, M, and H) components.
        The other hyperparameters are kept fixed: $\lambda$ for field smoothness regularization, $\gamma$ for rigid loss, and $\tau$ for valley loss threshold.}
        \centering
        \begin{tabular}{ *{3}{l} *{6}{c}}
            \toprule
            
            \multicolumn{1}{c}{Multiblur combination}
            &\multicolumn{1}{c}{}
            &\multicolumn{1}{c}{$\lambda$}
            &\multicolumn{4}{c}{$\omega$}
            &\multicolumn{1}{c}{$\gamma$}
            &\multicolumn{1}{c}{$\tau$}
            \\
            
            \cmidrule{4-9}
        
            \multicolumn{3}{c}{}
            &$FR$
            &S
            &M
            &H
            &\multicolumn{1}{c}{}
            &\multicolumn{1}{c}{}
            \\
            
            \midrule
            
            S-M-H
            &\multicolumn{1}{c}{}
            &$10^{-5}$
            &$0.4\phantom{0}$
            &$0.3\phantom{0}$
            &$0.2\phantom{0}$
            &$0.1\phantom{0}$
            &$0.01$
            &$32$
            \\
            
            M-H
            &\multicolumn{1}{c}{}
            &
            &$0.57$
            &$-$
            &$0.29$
            &$0.14$
            &
            &
            \\
        
            S-H
            &\multicolumn{1}{c}{}
            &
            &$0.5\phantom{0}$
            &$0.38$
            &$-$
            &$0.13$
            &
            &
            \\
            
            S-M
            &\multicolumn{1}{c}{}
            &
            &$0.45$
            &$0.33$
            &$0.22$
            &$-$
            &
            &
            \\
            
            H
            &\multicolumn{1}{c}{}
            &
            &$0.8\phantom{0}$
            &$-$
            &$-$
            &$0.2\phantom{0}$
            &
            &
            \\
            
            M
            &\multicolumn{1}{c}{}
            &
            &$0.67$
            &$-$
            &$0.33$
            &$-$
            &
            &
            \\
            
            S
            &\multicolumn{1}{c}{}
            &
            &$0.57$
            &$0.47$
            &$-$
            &$-$
            &
            &
            \\
    
            \bottomrule
        \end{tabular}
        \label{tab:params_multiblur}
        \end{table}

        \begin{table}[htbp]
        \caption{Performance comparison of multiblur schemes for FD-Net.
        PSNR and SSIM metrics are reported as mean (SD) across subjects. 
        Combinations over three different blur amounts are considered: small (S) , medium (M), and high (H) blur.
        Bold font denotes the best performing combination.
        The S-M-H multiblur combination is chosen as the multiresolution scheme for FD-Net due to its superior performance.}
        \centering
        \begin{tabular}{ *{1}{l} *{6}{c}}
            \toprule
            
            \multicolumn{1}{c}{Multiblur combination}
            &\multicolumn{1}{c}{}
            &\multicolumn{2}{c}{Image quality}
            &\multicolumn{1}{c}{}
            &\multicolumn{2}{c}{Field quality}\\
            
            \cmidrule{3-4} \cmidrule{6-7}
        
            \multicolumn{2}{l}{}
            &PSNR [dB]
            &SSIM [\%]
            &\multicolumn{1}{c}{}
            &PSNR [dB]
            &SSIM [\%]\\
            
            \midrule

            \textbf{S-M-H}
            &\multicolumn{1}{c}{}
            &$\mathbf{31.29\ (2.79)}$
            &$\mathbf{86.71\ (11.58)}$
            &\multicolumn{1}{c}{}
            &$\mathbf{22.48\ (5.69)}$
            &$\mathbf{83.09\ (10.37)}$\\
            
            M-H
            &\multicolumn{1}{c}{}
            &$31.26\ (2.87)$
            &$86.44\ (11.66)$
            &\multicolumn{1}{c}{}
            &$22.39\ (5.61)$
            &$82.79\ (10.52)$\\
            
            S-H
            &\multicolumn{1}{c}{}
            &$31.03\ (2.79)$
            &$85.98\ (11.75)$
            &\multicolumn{1}{c}{}
            &$21.60\ (5.91)$
            &$81.94\ (10.94)$\\
            
            S-M
            &\multicolumn{1}{c}{}
            &$31.01\ (2.73)$
            &$85.64\ (11.61)$
            &\multicolumn{1}{c}{}
            &$21.78\ (5.84)$
            &$82.14\ (10.84)$\\
            
            H
            &\multicolumn{1}{c}{}
            &$30.96\ (2.87)$
            &$85.83\ (11.73)$
            &\multicolumn{1}{c}{}
            &$21.52\ (5.75)$
            &$81.51\ (11.19)$\\
            
            M
            &\multicolumn{1}{c}{}
            &$29.01\ (2.84)$
            &$81.48\ (11.74)$
            &\multicolumn{1}{c}{}
            &$16.49\ (6.94)$
            &$74.29\ (14.91)$\\
            
            S
            &\multicolumn{1}{c}{}
            &$30.73\ (2.68)$
            &$84.80\ (11.56)$
            &\multicolumn{1}{c}{}
            &$21.12\ (5.99)$
            &$80.80\ (11.60)$\\
            
            \bottomrule
        \end{tabular}
        \label{tab:results_multiblur}
        \end{table}
        
        \subsubsection{Loss Ablation Study for FD-Net}\label{subsubsec:loss_ablation_study}

        An ablation study was conducted by removing one loss term at a time from \equationautorefname~\eqref{eqn:fdnet_maineqn} to investigate its contribution to the overall performance.
        Additionally, a version using only the MSE loss term (i.e., $\mathcal{L}_{MSE}$) was provided for reference.
        The results provided in \tableautorefname~\ref{tab:results_loss} indicate that the proposed FD-Net provides the best overall performance.
        Removal of the rigid loss slightly decreases the predicted field quality, while removal of the valley loss slightly decreases the predicted image quality.
        Removal of the bending energy loss has the most detrimental effect on  performance, leading to a significant drop in PSNR and SSIM down to the level of the MSE-only case.
        The proposed FD-Net improves the image quality by $0.27\text{dB}$ PSNR/$1.11\%$ SSIM and the field quality by $1.21\text{dB}$ PSNR/$1.74\%$ SSIM over the MSE-only case.

        \begin{table}[htbp]
        \caption{Performance comparison results for the loss ablation study for FD-Net.
        PSNR and SSIM metrics are reported as mean (SD) across subjects.
        Removal of a loss component is indicated by ``$\setminus$'' symbol followed by the removed loss term in curly braces.
        The full version of the loss is chosen for FD-Net as it provides the best overall performance.}
        \centering
        \begin{tabular}{ *{1}{l} *{6}{c}}
            \toprule
            
            \multicolumn{1}{c}{Loss terms}
            &\multicolumn{1}{c}{}
            &\multicolumn{2}{c}{Image quality}
            &\multicolumn{1}{c}{}
            &\multicolumn{2}{c}{Field quality}\\
            
            \cmidrule{3-4} \cmidrule{6-7}
        
            \multicolumn{2}{l}{}
            &PSNR [dB]
            &SSIM [\%]
            &\multicolumn{1}{c}{}
            &PSNR [dB]
            &SSIM [\%]\\
            
            \midrule
    
            $\mathcal{L}_{FD-Net}$
            &\multicolumn{1}{c}{}
            &$31.29\ (2.79)$
            &$86.71\ (11.58)$
            &\multicolumn{1}{c}{}
            &$22.48\ (5.69)$
            &$83.09\ (10.37)$\\
        
            $\mathcal{L}_{FD-Net} \setminus \left\{\mathcal{L}_{rigid}\right\}$
            &\multicolumn{1}{c}{}
            &$31.37\ (2.93)$
            &$86.65\ (11.68)$
            &\multicolumn{1}{c}{}
            &$22.21\ (5.86)$
            &$83.02\ (10.37)$\\
            
            $\mathcal{L}_{FD-Net} \setminus \left\{\mathcal{L}_{valley}\right\}$
            &\multicolumn{1}{c}{}
            &$31.23\ (2.79)$
            &$86.49\ (11.63)$
            &\multicolumn{1}{c}{}
            &$22.36\ (5.77)$
            &$83.10\ (10.27)$\\
            
            $\mathcal{L}_{FD-Net} \setminus \left\{\mathcal{L}_{BE}\right\}$
            &\multicolumn{1}{c}{}
            &$30.92\ (2.72)$
            &$85.51\ (11.67)$
            &\multicolumn{1}{c}{}
            &$21.46\ (5.82)$
            &$81.47\ (11.19)$\\
            
            $\mathcal{L}_{MSE}$
            &\multicolumn{1}{c}{}
            &$31.02\ (2.84)$
            &$85.50\ (11.70)$
            &\multicolumn{1}{c}{}
            &$21.37\ (5.92)$
            &$81.35\ (11.29)$\\
            
            \bottomrule
        \end{tabular}
        \label{tab:results_loss}
        \end{table}
        
    \subsection{Comparison with Competing Methods}\label{subsec:methods_comparison}
            
        Comprehensive quantitative evaluations and visual assessments of the proposed FD-Net and the competing methods were conducted with respect to the reference TOPUP results.

\begin{figure}[htbp]
        \centering
        \includegraphics[width=0.5\textwidth]{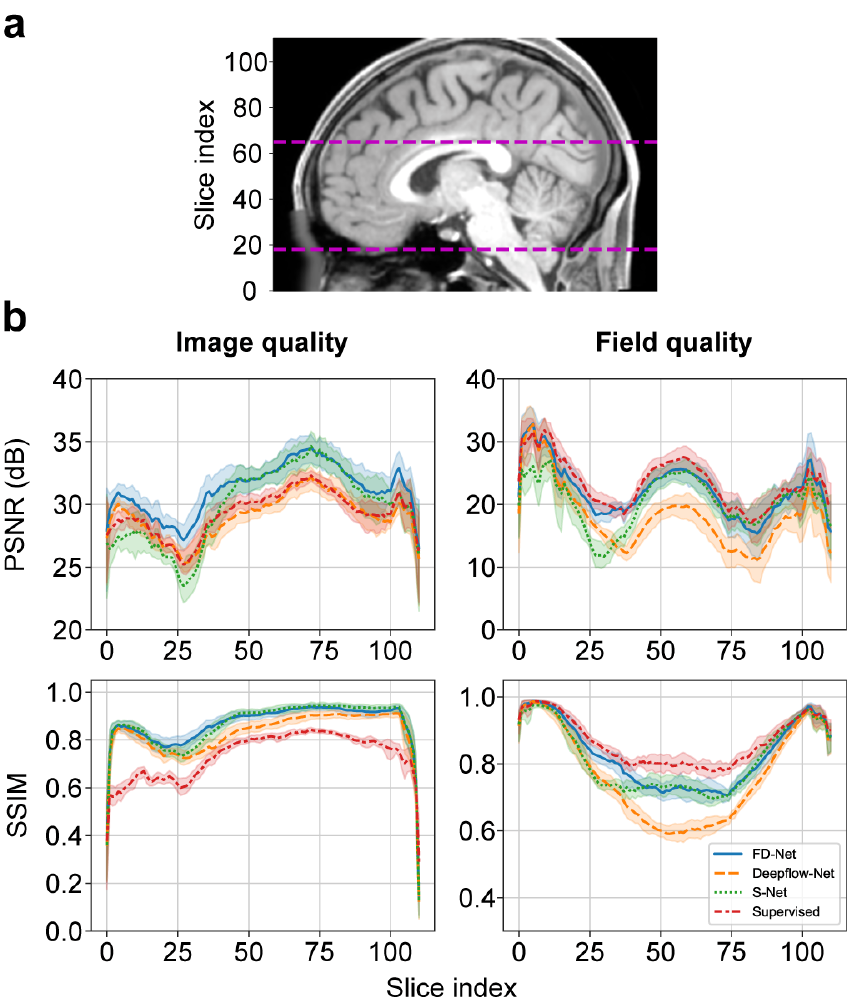}
        \caption{Slice-wise performance comparison of FD-Net and competing methods.
        (a) An example T\textsubscript{1} weighted image registered to the b0 volume of an individual subject to illustrate the anatomical locations corresponding to the slice indices.
        Magenta dashed lines indicate the locations of more challenging (lower line) and less challenging (upper line) slices in terms of distortion correction (see visual results in \figureautorefname~\ref{fig:visual_lower} and \figureautorefname~\ref{fig:visual_upper}).
        (b) PSNR (top row) and SSIM (bottom row) metrics for predicted image (left column) and predicted field (right column).
        Results are shown for FD-Net and competing methods as a function of slice index.
        For each method, the mean metric is shown along with the $\mathrm{95\%}$ confidence interval.}
        \label{fig:slicewise}
        \end{figure}

        \textit{Slice-Wise Evaluations:} \figureautorefname~\ref{fig:slicewise} demonstrates the performance of each method across different slices of the dataset.
        Since the dataset captured the same anatomy at the same orientation for all subjects, a given slice number corresponds to approximately the same anatomical location in all subjects.
        Hence, no additional intersubject registration was conducted for this analysis.
        The underlying anatomy is illustrated in \figureautorefname~\ref{fig:slicewise}a for a particular subject, where the T\textsubscript{1} weighted volume was registered to the corresponding b0 volume for display purposes, using FSL's FLIRT~\cite{Jenkinson_2001, Jenkinson_2002}.
        The results in \figureautorefname~\ref{fig:slicewise}b show that all methods have dips/peaks in performance at the same slice indices, providing insight into which slices are more/less challenging in terms of distortion correction.
        FD-Net outperforms all competing methods in terms of the predicted image quality, especially at the problematic lower brain slices where severe distortions are present.
        Moreover, the predicted field quality from FD-Net exceeds the competing methods, except for the supervised baseline.
        It should be noted that while the supervised baseline is able to match the TOPUP field better, it performs the worst in terms of predicted image quality.

\begin{figure}[htbp]
        \centering
        \includegraphics[width=0.5\textwidth]{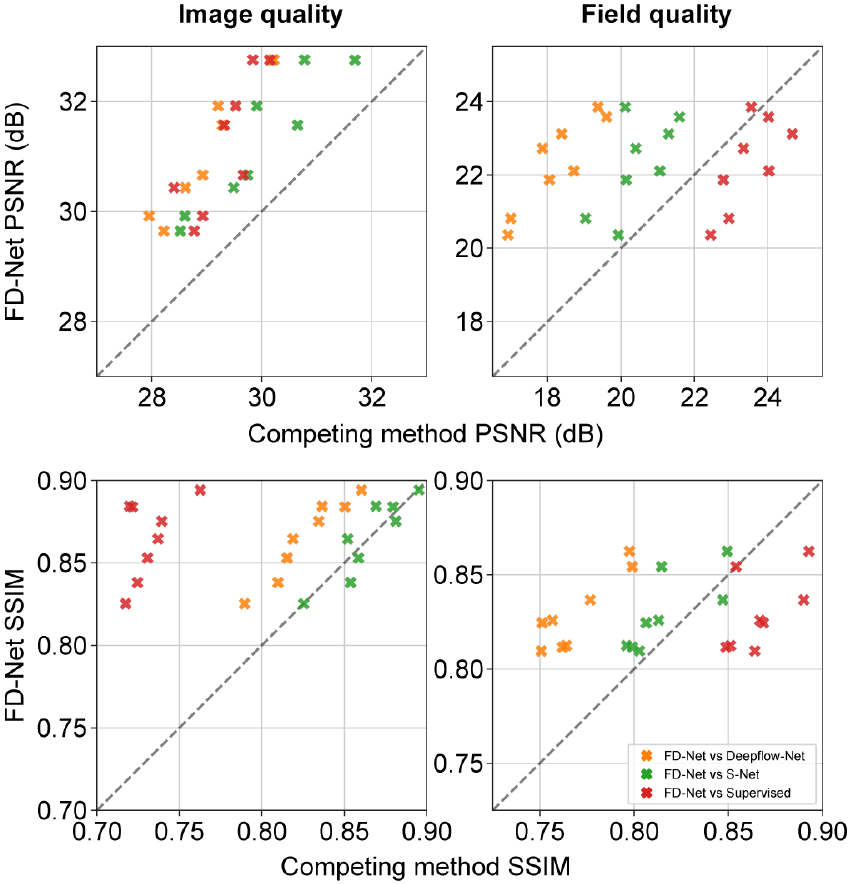}
        \caption{Subject-wise performance comparison of FD-Net against competing methods.
        PSNR (top row) and SSIM (bottom row) metrics for predicted image (left column) and predicted field (right column).
        Metrics are averaged across slices within individual subjects, and the mean metrics for the 8 test subjects are displayed as scatter plots.
        The vertical axis denotes FD-Net performance, whereas the horizontal axis denotes competing method performance (see legend).
        The results above the dashed identity lines indicate superior performance by FD-Net.}
        \label{fig:subjectwise}
        \end{figure}

        \textit{Subject-Wise Evaluations:} The performance of each method was assessed over all slices in the volume of a given subject, for each of the 8 subjects reserved for testing.
        \figureautorefname~\ref{fig:subjectwise} gives the scatter plots of mean PSNR and mean SSIM of FD-Net vs. each competing method for each subject, for a direct one-to-one performance comparison.
        In terms of image quality, FD-Net dominates over the competing methods, including the supervised baseline.
        While S-Net matches FD-Net in terms of SSIM over the predicted image quality, it lags behind in terms of PSNR.
        As for the predicted field quality, FD-Net is second only to the supervised baseline which was trained to directly fit the results from TOPUP.

        \textit{Overall Performance Evaluations:} The quantitative results in \tableautorefname~\ref{tab:results_comp} summarize the overall performance of each method across all subjects.
        FD-Net boosts image quality by $2.21\text{dB}$ PSNR/$4.01\%$ SSIM when compared to Deepflow-Net and $1.37\text{dB}$ PSNR/$0.27\%$ SSIM when compared S-Net.
        Field quality is boosted by  $4.24\text{dB}$ PSNR/$6.11\%$ SSIM when compared to Deepflow-Net and $2.03\text{dB}$ PSNR/$1.49\%$ SSIM when compared to S-Net.
        Compared to the supervised baseline, FD-Net largely boosts performance in terms of image quality by $1.97\text{dB}$ PSNR/$13.54\%$ SSIM, with a cost in field quality by $1.00\text{dB}$ PSNR/$3.61\%$ SSIM.

        \begin{table}[htbp]
        \caption{Performance comparison of FD-Net and the competing methods.
        PSNR and SSIM metrics are reported as mean (SD) across subjects.
        Bold font denotes the best performing method.}
        \centering
        \begin{tabular}{ *{1}{l} *{6}{c}}
            \toprule
            
            \multicolumn{1}{c}{Methods}
            &\multicolumn{1}{c}{}
            &\multicolumn{2}{c}{Image quality}
            &\multicolumn{1}{c}{}
            &\multicolumn{2}{c}{Field quality}\\
            
            \cmidrule{3-4} \cmidrule{6-7}
        
            \multicolumn{2}{l}{}
            &PSNR [dB]
            &SSIM [\%]
            &\multicolumn{1}{c}{}
            &PSNR [dB]
            &SSIM [\%]\\
            
            \midrule
        
            \textbf{Proposed FD-Net}
            &\multicolumn{1}{c}{}
            &$\mathbf{31.29\ (2.79)}$
            &$\mathbf{86.71\ (11.58)}$
            &\multicolumn{1}{c}{}
            &$22.48\ (5.69)$
            &$83.09\ (10.37)$\\
            
            Deepflow-Net
            &\multicolumn{1}{c}{}
            &$29.08\ (2.33)$
            &$82.70\ (11.72)$
            &\multicolumn{1}{c}{}
            &$18.24\ (6.48)$
            &$76.98\ (14.19)$\\
            
            S-Net
            &\multicolumn{1}{c}{}
            &$29.92\ (3.63)$
            &$86.44\ (12.16)$
            &\multicolumn{1}{c}{}
            &$20.45\ (5.43)$
            &$81.60\ (10.50)$\\
            
            Supervised baseline
            &\multicolumn{1}{c}{}
            &$29.32\ (2.24)$
            &$73.17\ (11.26)$
            &\multicolumn{1}{c}{}
            &$\mathbf{23.48\ (5.06)}$
            &$\mathbf{86.70\ (7.63)}$\\
            
            \bottomrule
        \end{tabular}
        \label{tab:results_comp}
        \end{table}
        
\begin{figure}[htbp]
        \centering
        \includegraphics[width=0.5\textwidth]{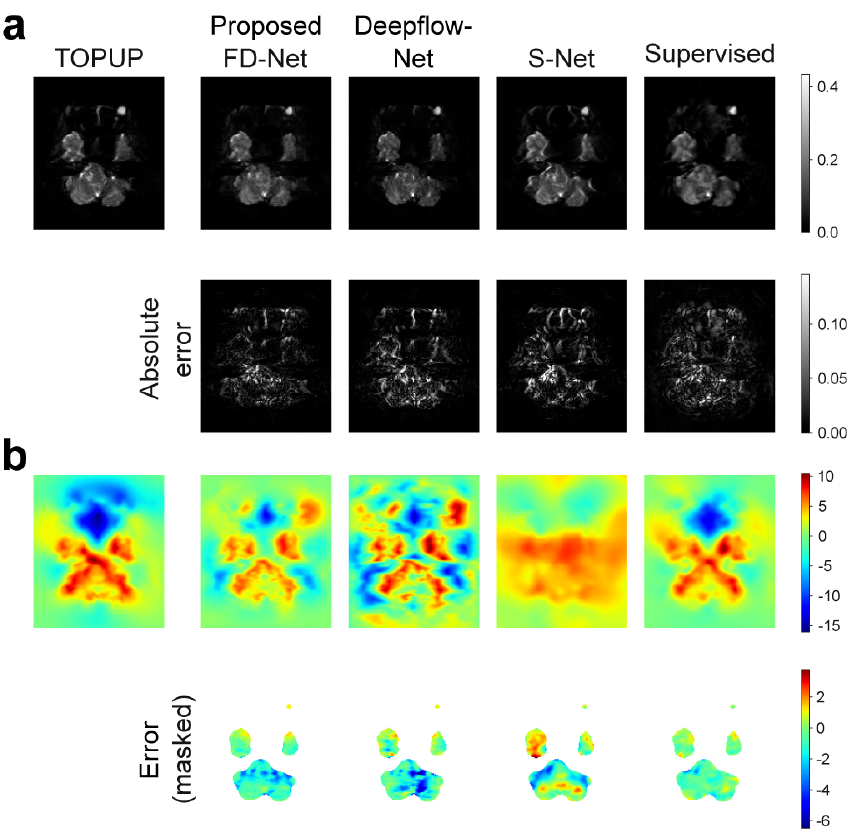}
        \caption{Visual results for FD-Net and competing methods from a lower brain slice, corresponding to a more challenging location in terms of distortion correction.
        TOPUP results are taken as reference.
        (a) Predicted images and absolute error maps with respect to TOPUP.
        The error maps are scaled by $1.25\times$ to a visibly discernible display window.
        (b) Predicted fields and the masked error maps with respect to TOPUP.
        The error maps were masked via a median Otsu threshold over the TOPUP image to remove the background regions.
        See the lower magenta dashed line in \figureautorefname~\ref{fig:slicewise}a for the anatomical location of this slice.}
        \label{fig:visual_lower}
        \end{figure}

\begin{figure}[htbp]
        \centering
        \includegraphics[width=0.5\textwidth]{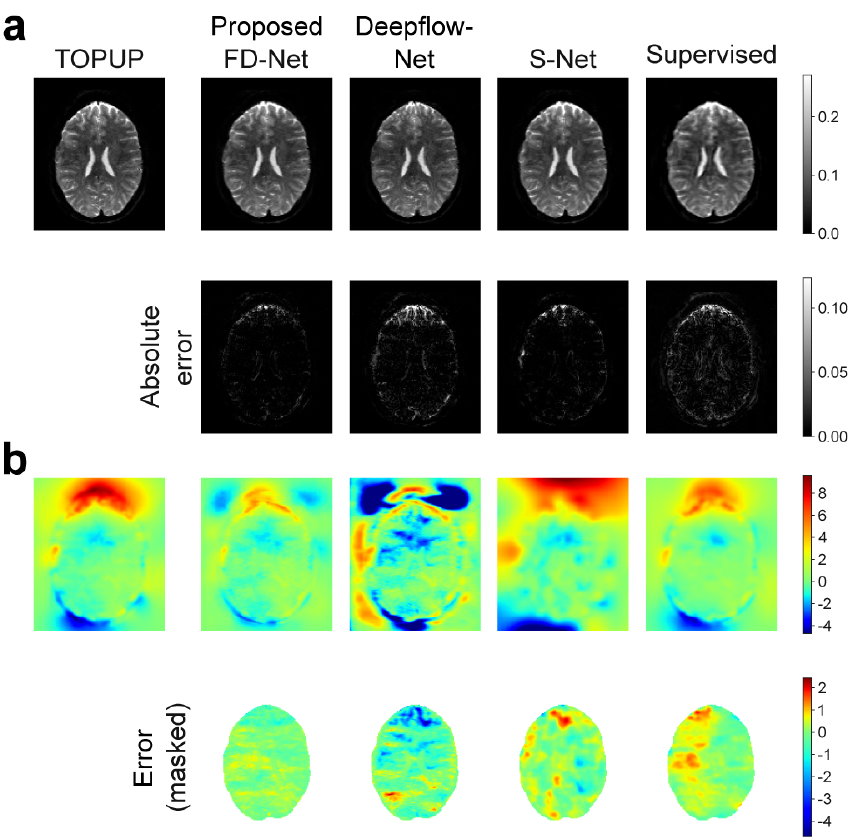}
        \caption{Visual results for FD-Net and competing methods from an upper brain slice, corresponding to a less challenging location in terms of distortion correction.
        TOPUP results are takes as reference.
        (a) Predicted images and absolute error maps with respect to TOPUP.
        The error maps are scaled by $1.25\times$ to a visibly discernible display window.
        (b) Predicted fields and the masked error maps with respect to TOPUP.
        The error maps were masked via a median Otsu threshold over the TOPUP image to remove the background regions.
        See the upper magenta dashed line in \figureautorefname~\ref{fig:slicewise}a for the anatomical location of this slice.}
        \label{fig:visual_upper}
        \end{figure}

\begin{figure}[htbp]
        \centering
        \includegraphics[width=0.5\textwidth]{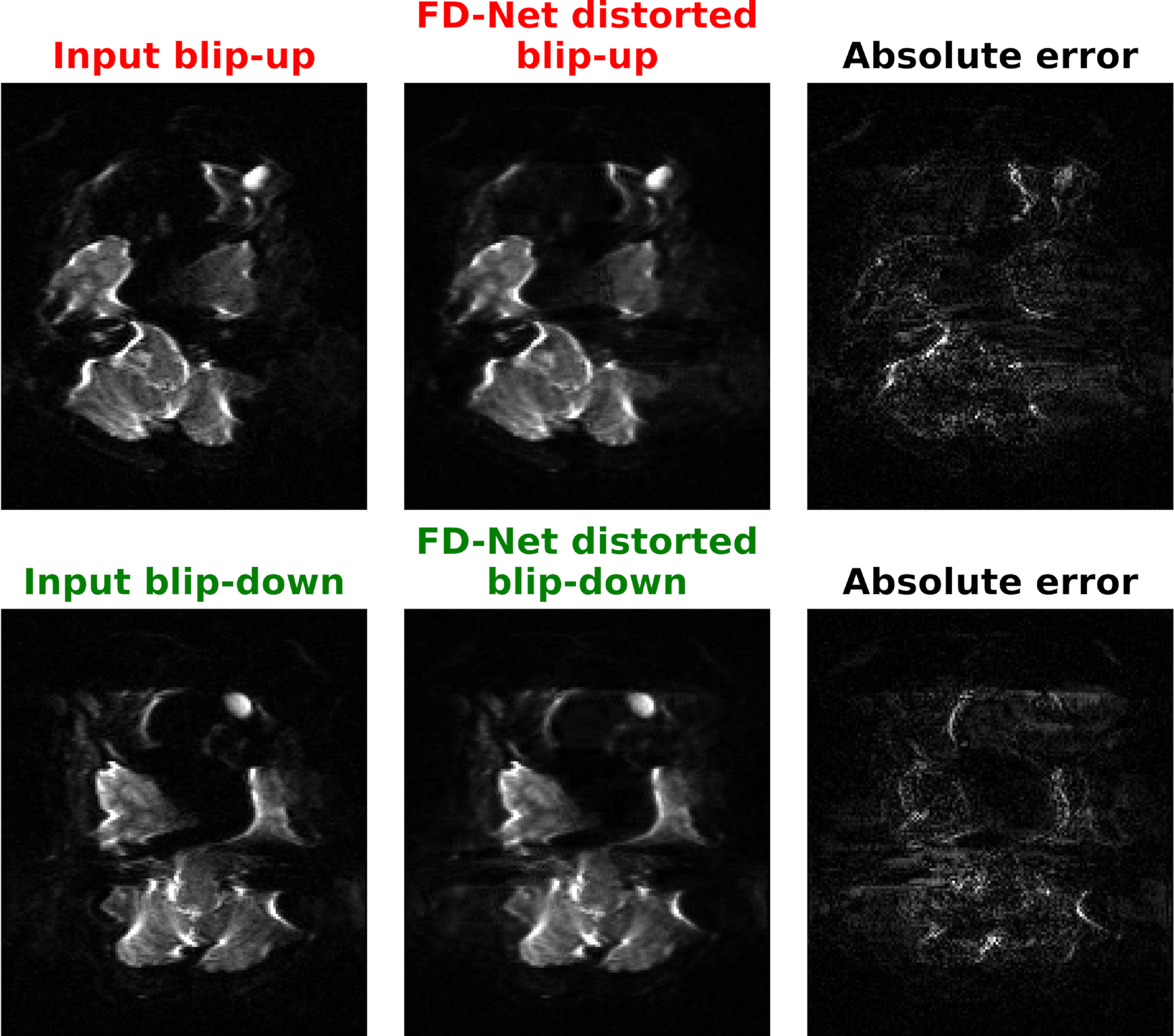}
        \caption{Visual results for the forward-distorted images from FD-Net for a lower brain slice.
        The input blip-up and blip-down EPI images are compared with the results of forward-distortion in FD-Net.
        The error maps are scaled by $2.5\times$ to a visibly discernible display window. 
        See the lower magenta dashed line in \figureautorefname~\ref{fig:slicewise} for the anatomical location of this slice.}
        \label{fig:visual_lower_fd}
        \end{figure}

\begin{figure}[htbp]
        \centering
        \includegraphics[width=0.5\textwidth]{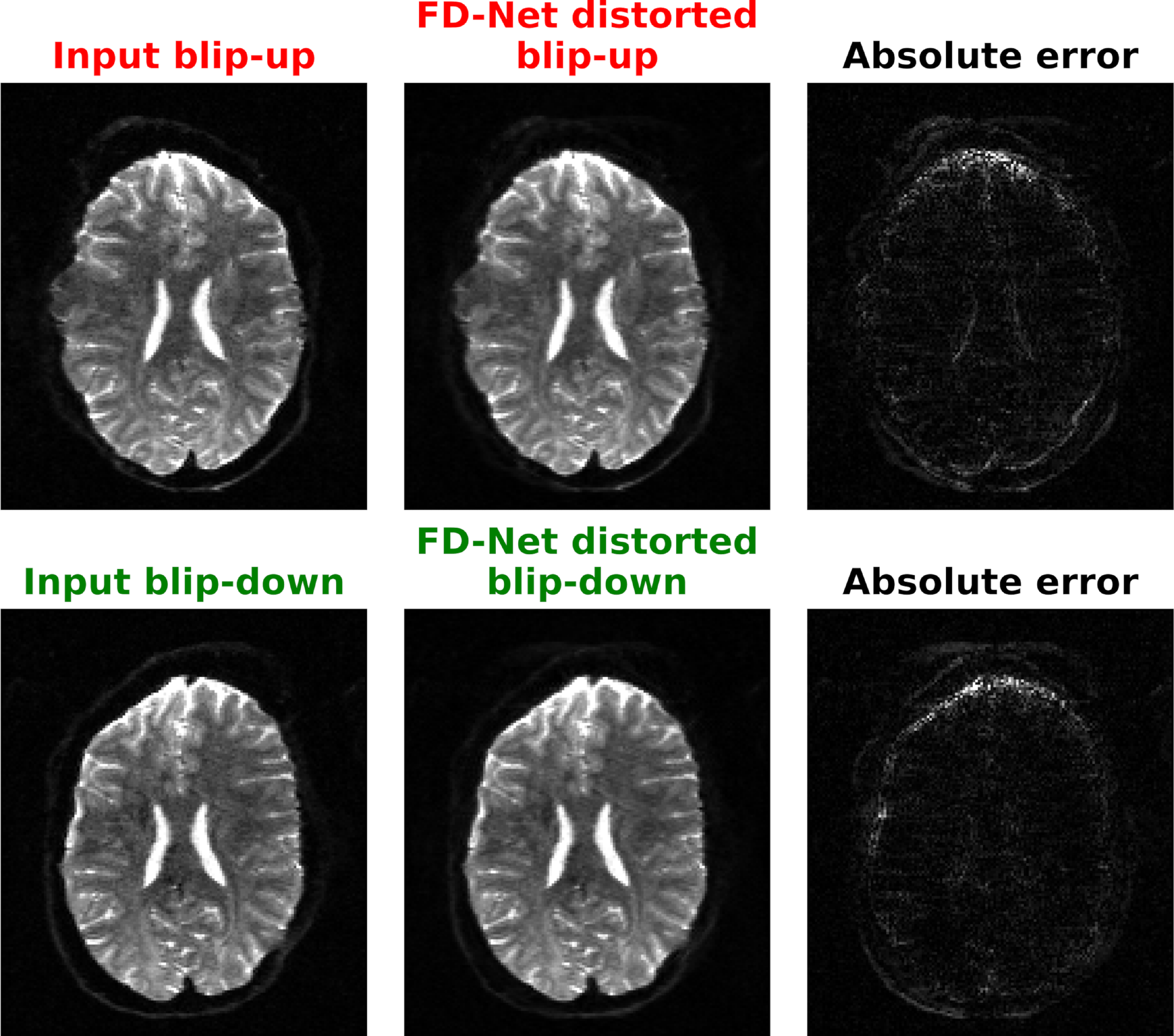}
        \caption{Visual results for the forward-distorted images from FD-Net for an upper brain slice.
        The input blip-up and blip-down EPI images are compared with the results of forward-distortion in FD-Net.
        The error maps are scaled by $2.5\times$ to a visibly discernible display window. See the upper magenta dashed line in \figureautorefname~\ref{fig:slicewise} for the anatomical location of this slice.}
        \label{fig:visual_upper_fd}
        \end{figure}

        \textit{Visual Assessments:} To visually compare the qualities of the predicted images and the predicted fields, example results from the slices marked in \figureautorefname~\ref{fig:slicewise}a are provided in \figureautorefname~\ref{fig:visual_lower} for a lower brain slice and \figureautorefname~\ref{fig:visual_upper} for an upper brain slice.
        These slices were chosen to represent the most and least challenging slices, corresponding to the dip and peak in PSNR in \figureautorefname~\ref{fig:slicewise}b, respectively.
        The error maps, as well as visual inspection of the predicted image and predicted field, indicate that FD-Net outperforms the other methods.
        This is especially true at the problematic lower brain slice example shown in \figureautorefname~\ref{fig:visual_lower}, where large distortions are present.
        The upper brain slice example in \figureautorefname~\ref{fig:visual_upper} exhibits distortions that are not as severe, indicating a less challenging problem for all methods to solve.
        For both cases, the predicted images from FD-Net have higher overall similarity to the TOPUP corrected image, with less artifacts present than the other methods.
        The field results also demonstrate that FD-Net produces the highest fidelity field, with smoothness and details preserved in a coherent manner.
        Additionally, the forward-distorted images generated by FD-Net closely match the input distorted images for both the lower and upper brain slices, as shown in \figureautorefname~\ref{fig:visual_lower_fd} and \figureautorefname~\ref{fig:visual_upper_fd}, respectively.

\section{Discussion}\label{sec:discussion}

In this work, we have proposed a deep forward-distortion model for unsupervised correction of susceptibility artifacts in EPI.
FD-Net is based on a multiresolution network model that estimates a single anatomically correct image along with a displacement field, given a pair of reversed-PE acquisitions.
Unsupervised learning is achieved by forward-distorting the anatomically correct image with the field, and enforcing consistency of the forward-distorted estimates to the input BU/BD acquisitions.
Our results indicate that this forward-distortion approach improves estimation fidelity for both the corrected image and field across a broad range of cross sections in the brain.
FD-Net outperforms competing unsupervised methods in image and field quality.
It also achieves higher image quality than the supervised baseline, while maintaining the field quality.

Unwarping-based methods rely on similarity losses between corrected BU/BD images to enable unsupervised learning.
As these losses are expressed in an inaccessible domain for which no explicit measurements are available, the resultant models can perform suboptimally under large displacements or intensity mismatches.
In particular, S-Net uses LCC between corrected images.
As a cross-modal similarity measure, LCC is known to be tolerant against intensity mismatches~\cite{Avants_2008}, but places higher emphasis on global features that can incur spatial blur in field estimates.
In turn, overly smooth field estimates and lack of density compensation in S-Net can limit its performance in regions of large displacements with abrupt susceptibility changes, particularly near the sinuses and ear canals.
To improve reliability against large displacements, Deepflow-Net performs density compensation by estimating pileups via linear interpolation of the grid point density map~\cite{Zahneisen_2020}.
However, the MSE loss that it adopts to measure similarity between corrected BU/BD images can lower tolerance against intensity mismatches and induce spatial blur in image estimates.
In contrast to unwarping-based methods, the proposed FD-Net leverages a forward-distortion approach based on the K-matrix formulation where density compensation is not needed.
For unsupervised learning, it uniquely measures the similarity between forward-distorted images, emulated from estimates of the anatomically-correct image and the field, and acquired BU/BD images.
As such, the similarity loss is expressed in the actual measurement domain, which can improve performance and reliability of FD-Net as suggested by our experimental results.
Quantitative assessments on field quality indicate that the supervised baseline provides a closer match to the TOPUP-estimated displacement field than FD-Net.
Yet, the apparent differences are relatively modest based on visual comparisons.
On the other hand, FD-Net achieves a notable boost in image quality over the supervised baseline, which is best attributed to the physics-based forward-distortion approach in FD-Net contributing to generalization performance~\cite{Aggarwal_2020}.

Here, we implemented all unsupervised correction methods by including a rigid loss for consistent and fair comparisons with FD-Net.
Based on \tableautorefname~\ref{tab:results_loss}, we observe that the rigid loss slightly influences image quality but achieves a modest boost in field quality.
This improvement can be attributed to the benefits of spatial registration to account for possible patient motion.
The empirical benefits of the rigid loss are expected to become more prominent for increasing levels of motion.
We also observe a modest improvement in image quality by inclusion of the valley loss.
This benefit can be attributed to the enhanced performance in regions of high field inhomogeneities by avoiding unrealistically large displacements.
Similarly, we observe that the bending energy loss that enforces field smoothness is critical to the performance of FD-Net.

As common in deep-learning methods, the trained weights of the FD-Net model are kept fixed during inference.
For models trained on limited datasets, this may results in suboptimal generalization to atypical anatomy.
In such cases, subject-specific optimization of model weights during inference might improve generalization at the expense of prolonged inference times~\cite{Narnhofer_2019, Korkmaz_2022}.
Here, modules within FD-Net were implemented based on convolutional backbones given their training efficiency.
To improve sensitivity to long-range context in brain images, self-attention based transformer backbones can be adopted~\cite{Dalmaz_2022}.
In the current study, all deep-learning models were effectively trained from scratch on relatively modest sized datasets including only 12 subjects.
In applications where training data are scarce, network modules can first be pre-trained on large public datasets, and later fine-tuned on the application-specific target datasets~\cite{Dar_2017}.
Lastly, here we assumed that only reversed-PE images are available as inputs to FD-Net.
In cases where additional measurements are viable to capture the field map and/or PSF, FD-Net could be modified to integrate these measurements for improved performance.

It is worth noting that the extent of susceptibility artifacts in EPI can also be reduced by modifying the imaging procedure.
For example, methods such as parallel imaging~\cite{Pruessmann_1999, Griswold_2002} or reduced FOV imaging~\cite{Saritas_2008, Barlas_2022} decrease sensitivity to field inhomogeneities by encoding a smaller FOV in the PE direction during data acquisition.
Similarly, multi-shot EPI~\cite{McKinnon_1993}, such as interleaved EPI~\cite{Nunes_2005}, can also be performed to reduce field sensitivity.
While powerful, these acquisition-based methods still require additional distortion correction in postprocessing.
The proposed FD-Net is compatible with this class of methods, as long as a reversed-PE acquisition is performed during imaging.

\section{Conclusions}\label{sec:conclusions}

In this work, we introduced a novel deep-learning approach for efficient and performant correction of susceptibility artifacts in EPI.
The proposed FD-Net estimates an anatomically correct image and a displacement field map.
It achieves unsupervised learning by leveraging a forward-distortion model to enforce consistency of the estimates to measured reversed-PE images.
FD-Net performs competitively with the reference TOPUP method, while offering notably faster inference as a deep-learning approach.
It also outperforms recent unsupervised correction methods that enforce similarity of unwarped reversed-PE images.
Therefore, FD-Net holds great promise for susceptibility-artifact correction in EPI applications.

\section*{Acknowledgments}
A preliminary version of this work was presented in the Annual Meeting of ISMRM in London, 2022.
This work was supported by the Scientific and Technological Council of Turkey (TÜBİTAK) via Grant 117E116.
Data were provided by the Human Connectome Project, WU-Minn Consortium (Principal Investigators: David Van Essen and Kamil Ugurbil; 1U54MH091657) funded by the 16 NIH Institutes and Centers that support the NIH Blueprint for Neuroscience Research; and by the McDonnell Center for Systems Neuroscience at Washington University.

\bibliographystyle{unsrt}  
\bibliography{references}

\end{document}